\documentclass[
preprint,
 amsmath,amssymb,
 aps,
prb,
floatfix,
]{revtex4-2}

\usepackage{graphicx}
\usepackage{dcolumn}
\usepackage{booktabs, tabularx, ragged2e}
\usepackage{bm}
\usepackage[mathlines]{lineno}
\usepackage{commath}
\usepackage{physics}
\usepackage{color}
\usepackage{mathtools}

\begin{document}

\preprint{APS/123-QED}

\title{Anisotropic Rashba splitting dominated by out-of-plane spin polarization in two-dimensional Janus $XA_{2}Y$ ($A$= Si, Sn, Ge; $X,Y$= Sb, Bi) with surface imperfection}

\author{Arif Lukmantoro}
\affiliation{Departement of Physics, Faculty of Mathematics and Natural Sciences, Universitas Gadjah Mada, Sekip Utara BLS 21 Yogyakarta 55186 Indonesia.}

\author{Moh. Adhib Ulil Absor}
\affiliation{Departement of Physics, Faculty of Mathematics and Natural Sciences, Universitas Gadjah Mada, Sekip Utara BLS 21 Yogyakarta 55186 Indonesia.}
\email{adib@ugm.ac.id}


\date{\today}

\begin{abstract}
The anisotropic Rashba effect allows for the manipulation of electron spins in a more precise and tunable manner since the magnitude of the Rashba splitting and orientation of the spin textures can be simply controlled by tuning the direction of the externally applied electric field. Herein, we predict the emergence of the anisotropic Rashba effect in the two-dimensional (2D) Janus $XA_{2}Y$ constructed from the group IV ($A$= Si, Sn, Ge) and group V ($X, Y$ = Sb, Bi) elements having trigonal prismatic structures but lacking in-plane mirror symmetry. Due to the lowering point group symmetry of the crystal, the Rashba spin splitting is enforced to become anisotropic around certain high symmetry points in the Brillouin zone and preserves the out-of-plane spin textures. We illustrate this behavior using density functional theory calculations supplemented with $\vec{k}\cdot\vec{p}$ analysis on the  Janus SbSi$_{2}$Bi monolayer as a representative example. Specifically, we observed large and anisotropic Rashba splitting with prominence contribution of the out-of-plane spin textures in the conduction band minimum around the $M$ point and valence band maximum around the $\Gamma$ point. More importantly, the anisotropic spin splitting and out-of-plane spin polarization are sensitively affected by surface imperfections, depending on the concentration and configuration of the $X$ and $Y$ elements in the 2D Janus $XA_{2}Y$ surface. Our study offers the possibility to realize the present systems for spintronics applications. 
\end{abstract}

\pacs{Valid PACS appear here}
\keywords{Suggested keywords}
\maketitle

\section{INTRODUCTION}

The spin-orbit coupling (SOC) in crystalline solids has gained increasing significance in spintronics since it offers a way to manipulate the spin of electrons without the need for an external magnetic field \cite{Manchon}. Several intriguing phenomena associated with SOC have been observed, such as spin relaxation \cite{Fabian, Averkiev}, spin Hall effect \cite{Qi}, spin galvanic effect \cite{Ganichev}, and spin ballistic transport \cite{Lu}. In systems lacking inversion symmetry, SOC generates a momentum-dependent spin-orbit field that breaks the spin degeneracy and creates non-trivial spin textures in the spin-split bands through the Rashba \cite{Rashba} and Dresselhaus \cite{Dress} effects. The Rashba effect has been extensively studied on systems with structural asymmetries, such as semiconductor quantum wells \cite{Stein, van, Nitta}, surfaces of heavy metals \cite{Gong, Varykhalov}, and various two-dimensional (2D) layered materials \cite{Singh, Adhib, Absor2023, WuKai, Yao2017}. In contrast, the Dresselhaus effect occurs in systems that exhibit bulk asymmetries, including bulk zincblende \cite{Soohyun} and wurtzite \cite{Wang2007} semiconductors. In particular, the Rashba effect has garnered much attention due to its ability to create non-equilibrium spin polarizations and electrically manipulate them by simply tuning the gate voltage \cite{Nitta, Kuhlen}. 

The Rashba SOC in its simplest form, which is linear in $k$, only produces isotropic spin splitting exhibiting concentric circles of Fermi contour lines with the in-plane chiral spin textures. However, when the anisotropy term is introduced in the Rashba SOC, the Fermi contour lines lose their circular shape and result in anisotropic spin splitting. Notably, the anisotropic Rashba effect has the potential to alter the characteristics of spin textures, such as giving rise to out-of-plane spin textures\cite{Xie, Absor2019, Sasmito, Absor2022, Absor2021}. Such spin textures could enable field-free spin–orbit torque switching of perpendicular magnetization \cite{Wang2023}, which may impact the transport properties such as the intrinsic spin-Hall effect \cite{Song2020} and anomalous Hall effect \cite{Yoo2021}. On the other hand, the anisotropic Rashba SOC, having different weights of Rashba SOC strength along the different wave vector directions, may induce unusual asymmetric spin relaxation \cite{Tao2021, Trushin}. Under the asymmetric spin relaxation processes, the presence of an electric current along with the Rashba SOC can induce current-induced spin polarization, which is very distinct from the spin Hall effect, known as the Rashba-Edelstein effect (REE) \cite{Edelstein, Sanchez2013, Isasa}, which allows for the conversion between charge and spin degrees of freedom. For the systems having strongly anisotropic Rasahba SOC, the largest REE occurs in the maximum accumulated spin density along the wave vector $\vec{k}$ direction with the largest Rashba SOC strength. Thus, it is possible to control precisely both the magnitude and orientation of the spin generated by the charge current by merely tuning the direction of the externally applied electric field. In addition, the anisotropic Rashba effect also significantly affects the spin Hall angle, as demonstrated by Yang et al. in graphene \cite{Yang16}. Therefore, by involving the anisotropy in the Rashba SOC and the spin textures, new opportunities can be explored in the field of spintronics.

The anisotropic Rashba effect can be achieved by incorporating the interplay between the Rashba and Dresselhauss SOCs realized through the interference of structure inversion asymmetry and bulk inversion asymmetry as previously demonstrated on various semiconductors quantum well \cite{Ganichev_2003, Giglberger}. However, due to the small of the Rashba SOC strength, efficient spin manipulation by an applied electric field is still questionable. Another approach to induce anisotropy involves reducing the crystal's symmetry, which can be accomplished through material engineering techniques such as surface reconstruction \cite{Simon, Miyamoto}, surface atomic or molecular absorption \cite{Friedrich, AbsIshii}, and dimensional reduction to 2D structures \cite{Popovi, Chakraborty, Zhang9, Sino, Hu}. The anisotropic Rashba splitting resulting from the presence of symmetry-reduced surface states has been previously reported on Au(110) \cite{Simon} and W(110) \cite{Miyamoto} surfaces which possess a point group symmetry of $C_{2v}$. Similarly, the absorption of a hydrogen atom on ZnO (10$\bar{1}$0) surface \cite{AbsIshii} and NH$_{3}$ and BH$_{3}$ molecules on BiAg$_{2}$/Bi(111) surface \cite{Friedrich}, both exhibiting a point group symmetry of $C_{s}$, demonstrates the generation of anisotropic Rashba effect via surface atomic or molecular absorption. Although surface reconstruction and surface molecular absorption have advantageous characteristics, they also bring forth challenges related to surface stability\cite{Guillope, Zakaryan2017}. Hence, utilizing material in the form of a 2D monolayer (ML) structure is the most suitable method to induce the anisotropic Rashba effect, which not only offers excellent stability but also provides a geometrical advantage in forming interfaces and heterostructures \cite{Castellanos, Gupta2021}, which holds the potential for miniaturization spintronic devices. However, to the best of our knowledge, only a few classes of the 2D ML systems have been reported to support the anisotropic Rashba effect including black-phosphorene \cite{Popovi}, BiTeI \cite{Zhang9}, and 2D Janus systems such as $M$Si$_{2}$P$_{x}$As$_{y}$ ($M$= Mo, W) MLs \cite{Rezavand} and transition metal dichalcogenides (TMDcs) $MXY$ ($M$= W, Mo, Pt; $X, Y$= S, Se, Te) MLs \cite{Sino, Hu, Chakraborty, Yagmurcukardes}. Hence, the search for a new 2D ML system that exhibits an anisotropic Rashba effect is greatly sought after, as it has the potential to expand the range of materials suitable for spintronics applications.

In this paper, through first-principles density-functional theory (DFT) calculations supplemented by the $\vec{k}\cdot\vec{p}$-based symmetry analysis, we predict the emergence of anisotropic Rashba effect having dominant out-of-plane spin polarization in the two-dimensional (2D) Janus $XA_{2}Y$ MLs constructed from the combination of the group IV ($A$= Si, Sn, Ge) and group V ($X, Y$= Sb, Bi) elements. These materials possess trigonal prismatic structures but lack in-plane mirror symmetry. Previously, the stability and electronic properties of various 2D group IV-V compounds have been widely studied \cite{LeeS2020, Ozdamar2018, Bafekry2021, Bafekry2020, Barreteau2016, Absor_2022}. The absence of the in-plane mirror symmetry in the Janus $XA_{2}Y$ MLs reduces the point group symmetry of the crystal, enforcing the Rashba SOC becomes anisotropic around certain high symmetry $\vec{k}$ points in the first Brillouin zone (FBZ) and preserves the significant out-of-plane spin textures. These properties are especially examined in the Janus SbSi$_{2}$Bi ML as a representative example of the Janus $XA_{2}Y$ MLs, which is notably apparent in the vicinity of the $M$ point in the conduction band minimum (CBM) and around the $\Gamma$ point in the valence band maximum (VBM). Importantly, the anisotropic spin splitting and out-of-plane spin textures are sensitively affected by surface imperfections, depending on the concentration and configuration of the $X$ and $Y$ elements in the 2D Janus $XA_{2}Y$ MLs surface. Our results demonstrate the potential to expand the range of experimentally accessible 2D materials, thus opening up new avenues for their utilization in spintronics applications.

\section{Computational Details}

We have carried out DFT calculations adopting norm-conserving pseudo-potentials and optimized pseudo-atomic localized basis functions \cite{Troullier} implemented in the OPENMX code \cite{OpenmX, Ozaki, Ozakikinoa, Ozakikinoa}. The generalized gradient approximation by Perdew, Burke, and Ernzerhof (GGA-PBE) \cite{gga_pbe, Kohn} was used as the exchange-correlation functional. We used the basis functions as the linear combination of multiple pseudo atomic orbitals (PAOs) generated using a confinement scheme \cite{Ozaki, Ozakikino, Ozakikinoa}. Here, two $s$-, two $p$-, two $d$-character numerical PAOs were applied. The FBZ integration was carried out using the $12\times12\times1$ $k$-point mesh. In order to prevent artificial interactions among the periodic images generated by the periodic boundary condition, we employed a periodic slab model for the Janus $XA_{2}Y$ MLs. This model included a sufficiently large vacuum layer (25 \AA) in the non-periodic direction. We optimized the lattice and positions of the atoms until the Hellmann-Feynman force components acting on each atom was less than $10^{-3}$ eV\AA\, where the energy convergence criterion was set to $10^{-9}$ eV. Phonon dispersion band is used to evaluate the dynamical stability of the Janus $XA_{2}Y$ MLs obtained by using ALAMODE code \cite{Tadano} based on the force constants obtained from the OpenMX code calculations. 

The spin vector component ($S_{x}$, $S_{y}$, $S_{z}$) of the spin polarization in the reciprocal lattice vector $\vec{k}$ was inferred by analyzing the spin density matrix\cite{Kotaka}. The spin density matrix, denoted as $P_{\sigma \sigma^{'}}(\vec{k},\mu)$, is computed using the spinor Bloch wave function, $\Psi^{\sigma}_{\mu}(\vec{r},\vec{k})$, through the following equation, 
\begin{equation}
\begin{aligned}
\label{1}
P_{\sigma \sigma^{'}}(\vec{k},\mu)=\int \Psi^{\sigma}_{\mu}(\vec{r},\vec{k})\Psi^{\sigma^{'}}_{\mu}(\vec{r},\vec{k}) d\vec{r}\\
                                  = \sum_{n}\sum_{i,j}[c^{*}_{\sigma\mu i}c_{\sigma^{'}\mu j}S_{i,j}]e^{\vec{R}_{n}\cdot\vec{k}},\\
\end{aligned}
\end{equation}
where $\Psi^{\sigma}_{\mu}(\vec{r},\vec{k})$ is obtained after self-consistent is achieved in the DFT calculation. In Eq. (\ref{1}), $S_{ij}$ is the overlap integral of the $i$-th and $j$-th localized orbitals, $c_{\sigma\mu i(j)}$ is expansion coefficient, $\sigma$ ($\sigma^{'}$) is the spin index ($\uparrow$ or $\downarrow$), $\mu$ is the band index, and $\vec{R}_{n}$ is the $n$-th lattice vector.  

In our DFT calculation, we considered the Janus $XA_{2}Y$ MLs where Si, Sn, and Ge atoms are chosen as $A$ (group IV) elements, while Sb and Bi atoms are taken as $X$ and $Y$ (group V) elements. These elements are chosen due to the larger atomic $Z$ number, which is expected to induce the significant SOC. To confirm the stability of Janus $XA_{2}Y$ MLs, we calculate the formation energy, $E_{f}$, by using the following relation,
\begin{equation}
\label{2}
E_{f}=E_{XA_{2}Y}-\frac{1}{n_{X}+n_{Y}+n_{A}}(n_{X}E_{X}+n_{Y}E_{Y}+n_{A}E_{A}),  
\end{equation}
where $E_{XA_{2}Y}$ is the total energy of Janus $XA_{2}Y$ MLs. $E_{X}$, $E_{Y}$, and $E_{A}$ are the chemical potential of
isolated $X$, $Y$, and $A$ atoms, respectively. $n_{X}$, $n_{Y}$, and $n_{A}$ are the number of $X$, $Y$, and $A$ atoms in the super cell or unit cell, respectively.

\begin{figure}
	\centering		
	\includegraphics[width=0.75\textwidth]{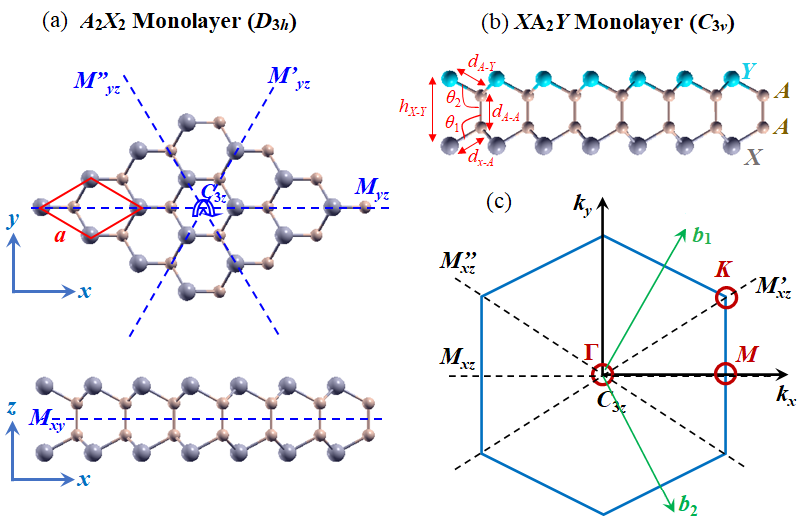}
	\caption{Atomic structures of (a) the $A_{2}X_{2}$ monolayers and (b) Janus $XA_{2}Y$ MLs corresponding to the first Brillouin zone (FBZ) (c) are shown. The unit cell of the crystal is indicated by the blue lines. The structural parameters including the bond length between $X$ and $A$ ($\delta_{X-A}$), the bond length between $Y$ and $A$ atoms ($\delta_{Y-A}$), the bond length between $A$ atoms ($\delta_{A-A}$), the out-of-plane distance between $X$ and $Y$ atoms ($h_{X-Y}$), and the bond angle of $X-A-A$ ($\theta_{1}$) and $A-A-Y$ ($\theta_{2}$) are indicated. The FBZ is characterized by the $\Gamma$, $\Gamma$, $M$, and $K$ high symmetry points. The symmetry operations in the crystal consisting of identity ($E$), in-plane mirror reflection ($M_{xy}$), out-of-plane mirror reflection ($M_{xz}$, $M_{xz}^{'}$, $M_{xz}^{"}$), three-fold rotation around the $z$-axis ($C_{3}$, $C^{2}_{3}$) are shown. }
	\label{figure:Figure1}
\end{figure}

\section{Results and Discussion}

First, we discuss the structural symmetry and stability of the 2D Janus $XA_{2}Y$ MLs. Previously, theoretical studies have been conducted on stable 2D systems formed by group IV-V compounds with an $A_{2}X_{2}$ ML structure \cite{LeeS2020, Ozdamar2018, Bafekry2021, Bafekry2020, Absor_2022}, and some of these systems have been successfully synthesized \cite{Barreteau2016}. These compounds consist of covalently bonded quadruple atomic layers arranged in an alternating $X-A-A-X$ sequence, resulting in a trigonal prismatic structure where $A$ atoms form a triangular prism around the $X$ dimer [Fig. 1(a)]. This structure is similar to that previously reported on GaSe ML \cite{Hirokazu}, which is reminiscent of the 2H-phase of TMDCs MLs \cite{Zhu2011}. The $A_{2}X_{2}$ monolayers possess crystal symmetry belonging to the $P\bar{6}m_{2}$ space group with a $D_{3h}$ point group. When these monolayers are transformed into Janus structures, namely $XA_{2}Y$ monolayers, by substituting group V atoms ($X$) on one side with different group V atoms, the in-plane mirror symmetry $M_{xy}$ is broken [see Fig. 1(b)], and hence the symmetry of the $A_{2}X_{2}$ monolayers becomes $C_{3v}$ point group. Under the $C_{3v}$ point group symmetry, the Janus $XA_{2}Y$ MLs exhibit three types of symmetry operations that preserve the structure: identity ($E$), out-of-plane mirror reflection ($M_{xz}$, $M_{xz}^{'}$, $M_{xz}^{"}$), and three-fold rotation around the $z$-axis ($C_{3}$, $C^{2}_{3}$). The corresponding FBZ is schematically shown in Fig. 1(c).

\begin{table}[h]
\caption{The optimized structural parameters of Janus $XA_{2}Y$ MLs including in-plane lattice constant $a$, the bond length between $X$ and $A$ ($\delta_{X-A}$), the bond length between $Y$ and $A$ atoms ($\delta_{Y-A}$), the bond length between $A$ atoms ($\delta_{A-A}$), and the out-of-plane distance between $X$ and $Y$ atoms ($h_{X-Y}$) are shown. All the structural parameters are measured in \AA. Bond angle of $X-A-A$ ($\theta_{1}$)  and $A-A-Y$ ($\theta_{2}$) are also indicated. The formation energy is represented by $E_{f}$, which is measured in eV.} 
\centering 
\begin{tabular}{ccc ccc ccc ccc ccc ccc ccc ccc ccc} 
\hline\hline 
 2D Janus &&& $a$ (\AA)  &&& $\delta_{X-A}$ (\AA)  &&& $\delta_{Y-A}$ (\AA) &&& $\delta_{A-A}$ (\AA)  &&& $h_{X-Y}$ (\AA)  &&& $\theta_{1}$ &&& $\theta_{2}$ &&& E$_{for}$ (eV) \\ 
\hline 
SbSi$_2$Bi     &&& 4.01 &&& 2.63 &&& 2.68  &&& 2.35 &&& 4.96 &&& 120.43   &&& 118.35 &&& -4.8\\
SbGe$_2$Bi     &&& 4.19 &&& 2.73 &&& 2.77  &&& 2.51 &&& 5.13 &&& 119.43   &&& 117.64 &&& -4.25 \\
SbSn$_2$Bi     &&& 4.65 &&& 2.98 &&& 3.01  &&& 2.90 &&& 5.55 &&& 116.75   &&& 115.76 &&& -3.16\\
Si$_2$Sb$_{2}$ &&& 4.00 &&& 2.67 &&& -     &&& 2.36 &&& 4.86 &&& 118.43   &&& -     &&& -    \\
Si$_2$Bi$_{2}$ &&& 4.13 &&& 2.72 &&& -     &&& 2.35 &&& 4.95 &&& 118.62   &&& -     &&& -    \\
Ge$_2$Sb$_{2}$ &&& 4.13 &&& 2.71 &&& -     &&& 2.51 &&& 5.10 &&& 118.44   &&& -     &&& -    \\
Ge$_2$Bi$_{2}$ &&& 4.27 &&& 2.80 &&& -     &&& 2.51 &&& 5.16 &&& 118.24   &&& -     &&& -    \\
Sn$_2$Sb$_{2}$ &&& 4.37 &&& 2.98 &&& -     &&& 2.89 &&& 5.71 &&& 119.16   &&& -     &&& -    \\
Sn$_2$Bi$_{2}$ &&& 4.51 &&& 2.97 &&& -     &&& 2.87 &&& 5.72 &&& 118.62   &&& -     &&& -    \\
\hline\hline 
\end{tabular}
\label{table:Table 1} 
\end{table}

\begin{figure}
	\centering		
	\includegraphics[width=1.0\textwidth]{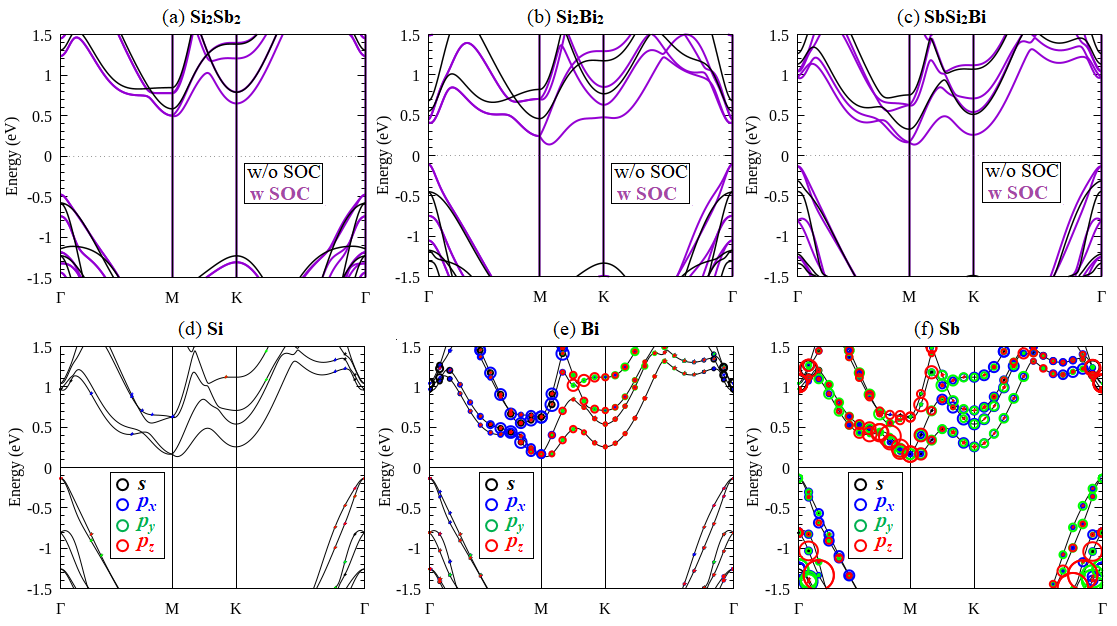}
	\caption{Electronic band structures of (a) pure Si$_{2}$Bi$_{2}$, (b) Janus SbSi$_{2}$Bi, and (c) Si$_{2}$Bi$_{2}$ MLs calculated without (black lines) and with (red lines) spin-orbit coupling. Orbital-resolved electronic band structures of Janus SbSi$_{2}$Bi ML calculated for: (d) Si, (e) Bi, and (f) Sb atoms are presented. Here, blue, green, and red colors of circle represent $s$, $(p_{x}+p_{y})$, and $p_{z}$ orbitals. The radii of the circles reflect the magnitudes of spectral weight of the particular orbitals to the band.}
	\label{figure:Figure2}
\end{figure}

The optimized structural-related parameters associated with Janus $XA_{2}Y$ MLs are presented in Table I. It is observed that the lattice constants of Janus $XA_{2}Y$ MLs decrease as the atomic number of $A$ elements decreases. Notably, SbSn$_{2}$Bi ML demonstrates the largest value of 4.65 \AA, while SbSi$_{2}$Bi ML exhibits the smallest value of 4.01 \AA. To provide a basis for comparison, we also conducted calculations for $A_{2}X_{2}$ monolayers and discovered that the lattice constants of Janus $XA_{2}Y$ MLs are approximately the average values between $A_{2}X_{2}$ and $A_{2}Y_{2}$ MLs. Furthermore, due to the strong covalent bond between $A$ atoms, the bond length between $A$ atoms ($\delta_{A-A}$) in Janus $XA_{2}Y$ MLs remains largely unchanged. However, as indicated in Table I, the bond length between $X$ ($Y$) and $A$ atoms, $\delta_{X-A}$ ($\delta_{Y-A}$), increase for Janus systems as the atomic number of the $A$ elements increases, reflecting the trend observed in lattice constants. Additionally, the variation in the distance between $X$ and $Y$ atoms ($h_{X-Y}$) in the out-of-plane direction, as well as the bond angle of $X-A-A$ ($\theta_{1}$) and $A-A-Y$ ($\theta_{2}$) follows the opposite pattern as the lattice constant trends. 

The stability of the 2D janus $XA_{2}Y$ MLs is confirmed by calculating the formation energy $E_{f}$. It is revealed that all the considered Janus $XA_{2}Y$ MLs exhibit negative formation energy, $E_{f}<0$ [see Table I], indicating that they are energetically favorable to be realized in experiments, similar to the 2D Janus $M$Si$_{2}$P$_{x}$As$_{y}$ MLs \cite{Rezavand} and Janus TMDCs $MXY$ MLs \cite{Sino, Hu, Chakraborty}. In addition, the stability of the 2D Janus $XA_{2}Y$ MLs is also confirmed by the calculated phonon dispersion bands shown in Figs. S1 in the supplementary materials \cite{Supporting}. It is clearly seen that there are no imaginary frequencies found in the phonon dispersion bands, showing that the optimized Janus $XA_{2}Y$ MLs is dynamically stable. Considering the fact that the Janus SbSi$_{2}$Bi ML exhibits the lowest formation energy [see Table I], in the following discussion we will particularly focus on the Janus SbSi$_{2}$Bi ML as a representative examples of the Janus $XA_{2}Y$ MLs.  

Figs. 2(a)-(c) depict the electronic band structures of the Janus SbSi$_{2}$Bi ML compared to those of the pure Si$_{2}X_{2}$ ($X$=Sb, Bi) MLs. The calculations were performed both without (black lines) and with (red lines) consideration of the SOC. Similar to the pure Si$_{2}X_{2}$ MLs, the Janus SbSi$_{2}$Bi ML exhibits an indirect band gap, where the CBM and VBM are situated at the $M$ and $\Gamma$ points, respectively. The calculated band gap of the Janus SbSi$_{2}$Bi ML is 0.70 eV at the GGA-PBE level, which is smaller than that of the pure Si$_{2}$Sb$_{2}$ (0.94 eV) and Si$_{2}$Bi$_{2}$ (1.2 eV) MLs. Analysis of the band projections onto the atoms confirms that the CBM of the Janus SbSi$_{2}$Bi ML predominantly arises from strong admixtures between Bi-$p_{x}+p_{y}$, Bi-$p_{z}$, Sb-$p_{x}+p_{y}$, and Sb-$p_{z}$ orbitals, while the VBM is mainly contributed by the Sb-$p_{y}$ and Sb-$p_{z}$ orbitals [Figs. 2(d)-2(f)].

\begin{figure*}
	\centering
		\includegraphics[width=1.0\textwidth]{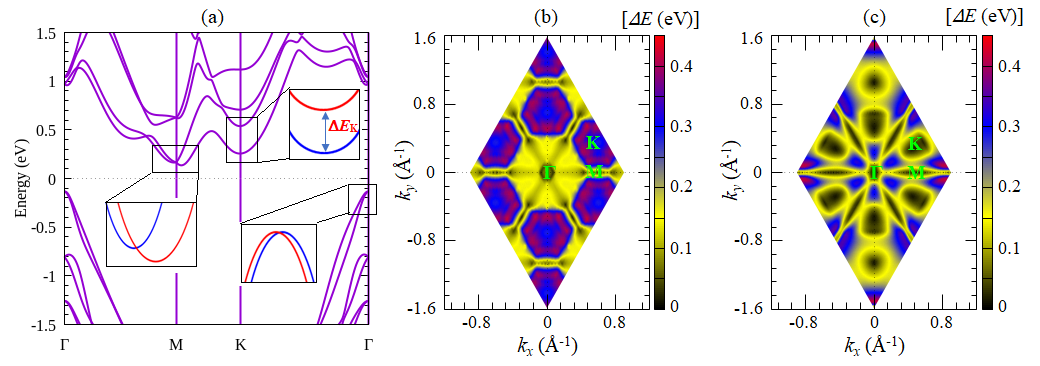}
	\caption{(a) The Highlight spin-split bands of the Janus SbSi$_{2}$Bi ML near the Fermi level around $M$ and $K$ point at the CBM and $\Gamma$ point at the VBM. (b)-(c) Spin-splitting energy of the Janus SbSi$_{2}$Bi monolayer mapped in the first Brillouin zone for CBM and VBM, respectively, are shown. The magnitude of the spin-splitting energy, $\Delta E$, defined as $\Delta E=|E(k,\uparrow)-E(k,\downarrow)|$, where $E(k,\uparrow)$ and $E(k,\downarrow)$ are the energy bands with up spin and down spin, respectively, is represented by the color scales. }
	\label{fig:Figure3}
\end{figure*}

Turning the SOC, the indirect band gap of the SbSi$_{2}$Bi ML remains but decreases by approximately 0.3 eV, comparable to the pure Si$_{2}$Bi$_{2}$ ML (0.25 eV) and much smaller than that of the pure Si$_{2}$Sb$_{2}$ ML (0.9 eV). Due to the absence of inversion symmetry in both the Janus SbSi$_{2}$Bi ML and pure Si$_{2}X_{2}$ MLs, all bands exhibit spin splitting, except at high symmetry points ($\Gamma$ and $M$) due to time reversibility [Figs. 2(a)-(c)]. However, for the pure Si$_{2}X_{2}$ MLs, the presence of in-plane mirror symmetry $M_{xy}$ [see Fig. 1(a)] preserves the spin degeneracy of bands along the $\Gamma-M$ line [Figs. 2(a)-2(b)]. This degeneracy is lifted in the Janus SbSi$_{2}$Bi ML [Fig. 2(c)] because of the broken $M_{xy}$ mirror symmetry [see Fig. 1(b)]. The observed spin splitting along the $\Gamma-M$ line in the Janus SbSi$_{2}$Bi ML is expected to be crucial in generating spin-polarized states through the Rashba effect, which holds significance for the operation of spin-field effect transistors (SFETs) \cite{Datta}.

\begin{figure*}
	\centering
		\includegraphics[width=1.0\textwidth]{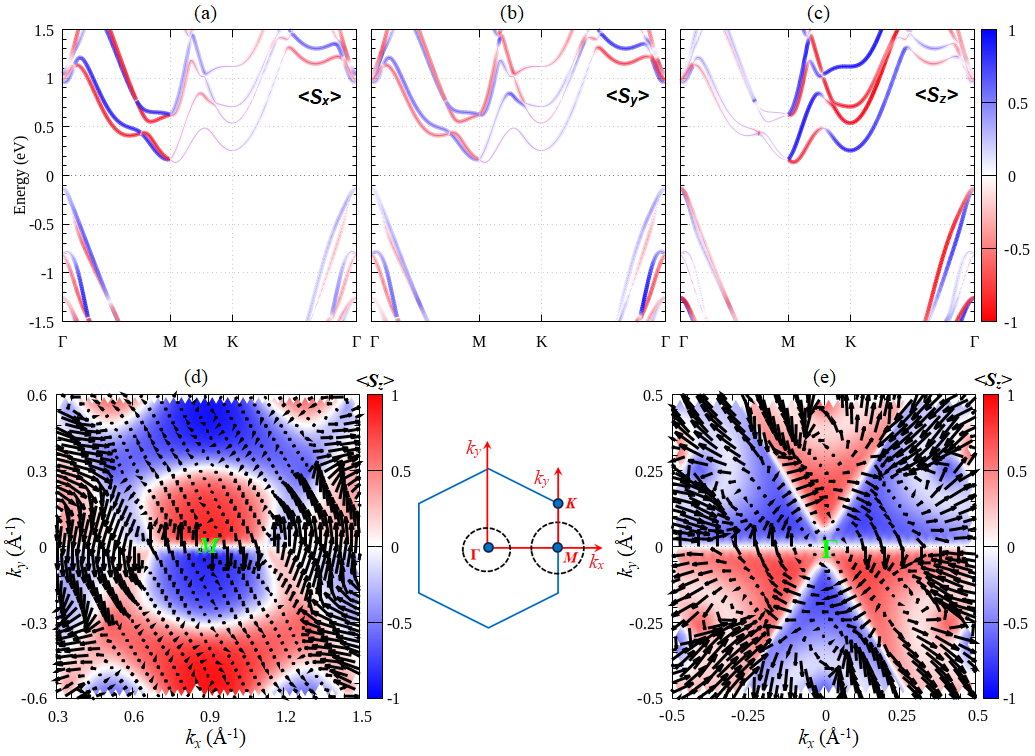}
	\caption{Spin-resolved projected to the bands of Janus SbSi$_{2}$Bi monolayer for: (a) $S_{x}$,  (b) $S_{y}$, and  (c) $S_{z}$ spin component of the spin polarization $\vec{S}$ is shown. The spin textures mapped in the FBZ calculated for the bands around (d) $M$ and (e) $\Gamma$ points are shown. The arrow represent the magnitude of the in-plane spin spin components ($S_{x}$, $S_{y}$), while the color represents the out-of-plane spin component ($S_{z}$). The insert show the position of the $M$ and $\Gamma$ points in the FBZ used in the spin textures calculations.}
	\label{fig:Figure4}
\end{figure*}

In order to further examine the spin-splitting characteristics within the Janus SbSi$_{2}$Bi ML, our focus was directed towards the spin-split bands near the Fermi level, as depicted in Figure 3(a). The outcomes demonstrate significant spin splitting at the CBM around the $M$ and $K$ points, as well as at the VBM around the $\Gamma$ point [as indicated in the inset of Figure 3(a)]. Notably, the spin splitting observed around the $M$ and $\Gamma$ points displays an anisotropic behavior, evident in the calculated spin-splitting energy mapped across the entire FBZ region, as shown in Figures 3(b)-3(c). In this regard, the spin-splitting energy exhibits maximum values along the $M-K$ line and minimum values along the $\Gamma-M$ line, both in the CBM and VBM. For instance, at the CBM, the most significant spin splitting, up to 0.65 eV, occurs at the $k$ point along the $M-K$ line, whereas only 0.1 eV is observed at the $k$ point along the $\Gamma-M$ line [Figure 3(b)]. In particularly, a substantial spin splitting of 0.52 eV has been identified at the non-time reversal invariant $K$ point in the CBM, which is notably larger than that observed in various 2D Janus MLs (0.14 - 0.48 eV) \cite{Rezavand, Sino, Hu, Chakraborty}. The considerable magnitude of spin splitting observed in these present systems signifies their potential for spintronics functionality, operating even at room temperature \cite{Yaji2010}.

To further analyze the Rashba anisotropy observed in the Janus SbSi$_{2}$Bi ML, we present in Figures 4(a)-(c) the expectation values of spin components ($S_{x}$, $S_{y}$, $S_{z}$) projected onto the spin-split bands near the Fermi level. Our findings reveal that, in addition to the in-plane spin components ($S_{x}$, $S_{y}$) observed in the spin-split polarized states at the $k$ point along the $\Gamma-M$ line, a significant contribution of the out-of-plane spin component ($S_{z}$) is identified in the spin-split polarized states at the $k$ point along the $M-K-\Gamma$ line. This anisotropy of the spin polarization direction in $k$-space within the Janus SbSi$_{2}$Bi ML is most clearly illustrated in Figures 4(d)-4(e) by the spin textures projected onto the FBZ around the $M$ and $\Gamma$ points in the CBM and VBM, respectively. In these figures, for clarity, we indicate the direction of the in-plane spin polarization using arrows, while the out-of-plane spin polarization is represented by colors. Around the $M$ point, the spin textures exhibit a $C_{s}$ symmetry of spin polarization, characterized by the reversal of the out-of-plane spin polarization mirrored at $k_{x}=0$ [Figure 4(d)]. Conversely, around the $\Gamma$ point in the VBM, the spin textures display a $C_{3v}$ symmetry of spin polarization, characterized by the in-plane spin polarization and three-fold pattern of the out-of-plane spin orientation [Figure 4(d)]. These particular spin textures observed around the $M$ and $\Gamma$ points deviate from the conventional Rashba spin textures commonly observed in semiconductor quantum wells \cite{Stein, van, Nitta}, surfaces of heavy metals \cite{Gong, Varykhalov}, and various two-dimensional (2D) layered materials \cite{Singh, Adhib, Absor2023, WuKai, Yao2017}. Notably, the anisotropy in the spin polarization direction, with a prominent out-of-plane spin polarization component, can impede electron back-scattering in spin transport and induce long-lived helical spin-wave modes by suppressing the spin-relaxation\cite{Dyakonov}. This unique feature holds great promise for efficient spintronics applications.  

To understand the origin of the ansiotropic of the spin splitting and spin polarization direction, we developed a $\vec{k}\cdot\vec{p}$ model based on the symmetry analysis. The band dispersion around the high symmetry point can be determined by the Hamiltonian $H(k)$ allowed by symmetry, so that $H(\vec{k})=O^{\dagger}H(\vec{k})O$, where $O$ represents symmetry operations associated with the wave vector group ($G$) corresponding to the high-symmetry point and time-reversal symmetry ($T$). The invariant $H(k)$ should obtain the following condition \cite{Winkler},
\begin{equation}
\label{3}
H_{G}(\vec{k})= D(O)H(O^{-1}\vec{k})D^{-1}(O),\ \forall O\in G, T  
\end{equation}
where $D(O)$ is the matrix representation of operation $O$ belonging to point group of the wave vector $G$. For our Janus systems, we focus on the high symmetry point of the FBZ depicted in Figure 1(c), and the corresponding little group of the wave vector is outlined in Table II. The corresponding transformations of $k$ and $\vec{\sigma}$ for all symmetry allowed terms are listed in Table III.

\begin{table*}
\caption{The little point group of the wave vector $\vec{k}$ in the high symmetry points and lines in the FBZ for Janus $XA_{2}Y$ MLs. The symmetry operations occured in the high symmetry points and lines in the FBZ are indicated in Fig. 1(c).} 
\begin{tabular}{ccc  ccc  ccc } 
\hline\hline 
  $\vec{k}$ point or line   &&&  Janus $XA_{2}Y$ ($C_{3v}$) &&&  symmetry operations \\ 
\hline 
$\Gamma$       &&&  $C_{3v}$  &&&     $E$, ($M_{xz}$, $M_{xz}^{'}$, $M_{xz}^{"}$), ($C_{3}$, $C^{2}_{3}$) \\
     $M$       &&&   $C_{s}$  &&&     $E$, $M_{xz}$ \\
     $K$       &&&  $C_{3}$   &&&     $E$, ($C_{3}$, $C^{2}_{3}$) \\
$\Gamma-M$     &&&  $C_{s}$   &&&     $E$, $M_{xz}$  \\
$\Gamma-K$     &&&  $C_{1}$   &&&     $E$   \\
		 $M-K$     &&&  $C_{1}$   &&&     $E$    \\		
\hline\hline 
\end{tabular}
\label{table:Table 2} 
\end{table*}

\begin{table*}
\caption{The transformations of ($\sigma_{x}$, $\sigma_{y}$, $\sigma_{z}$) and ($k_{x}$, $k_{y}$) with respect to the generators of $C_{3v}$, $C_{s}$, and $C_{1}$ point groups and time-reversal operator ($T$). Note that only these generators along with time-reversal $T=i\sigma_{y}K$ operation, where $K$ is complex-conjugation operator, are considered to construct the $\vec{k}\cdot\vec{p}$ model. The last column shows the terms which are invariant under point-group operation.} 
\begin{tabular}{c c c c} 
\hline\hline 
             \textbf{Operations}  & $(k_{x}, k_{y})$ & $(\sigma_{x}, \sigma_{y}, \sigma_{z})$ & \textbf{Invarian terms}  \\ 
\hline 
  $C_{3z}=e^{-i\pi/3\sigma_{z}}$ & $\frac{1}{2} \left([-k_{x}+\sqrt{3}k_{y}], [-\sqrt{3}k_{x}-k_{y}]\right)$  & $\left(\frac{1}{2}[-\sigma_{x}+\sqrt{3}\sigma_{y}], \frac{1}{2}[-\sqrt{3}\sigma_{x}-\sigma_{y}], \sigma_{z}\right)$         & $k_{x}\sigma_{y}-k_{y}\sigma_{x}$, \\
	         &        &          &  $k_{y}(3k_{x}^{2}-k_{y}^{2})\sigma_{z}$,  \\
					 &        &          &  $k_{x}(k_{x}^{2}-3k_{y}^{2})\sigma_{z}$  \\

$M_{xz}=i\sigma_{y}$   &  $(k_{x}, -k_{y})$      &   $(-\sigma_{x}, \sigma_{y}, -\sigma_{z})$       &  $k_{i}^{m}k_{y}\sigma_{x}$, $k_{i}^{m}k_{y}\sigma_{z}$, \\
           &        &          &  $k_{i}^{m}k_{x}\sigma_{y}$ \\
					 &        &          &  ($i=x,y$; $m=0,2$) \\
			
$T=i\sigma_{y}K$   &  $(-k_{x}, -k_{y})$      &   $(-\sigma_{x}, -\sigma_{y}, -\sigma_{z})$       &  $k_{i}\sigma_{j}$ \\
           &        &          &  ($i=x,y$; $j=x,y,z$)\\
          
\hline\hline 
\end{tabular}
\label{table:Table 3} 
\end{table*}

First, we derive an effective $\vec{k}\cdot\vec{p}$ Hamiltonian around the high symmetry $M$ point. Here, the little point group of the wave vector belongs to $C_{s}$ point group [see Table II]. This point group comprises mirror planes $M_{xz}$ in addition to the identity operation $E$. Consequently, both the wave vector $\vec{k}$ and the spin vector $\vec{\sigma}$ can be transformed according to the symmetry operation in the $C_{s}$ point group and time-reversal symmetry $T$. By collecting all terms that invariant under the symmetry operations of the $C_{s}$ point group [Table III], the following effective Hamiltonian for the ($k_{x}-k_{y}$) plane holds \cite{Vajna}:
\begin{equation}
\label{4}
H_{M}(k)= H_{0}(k)+ \alpha k_{x}\sigma_{y}+ \alpha^{'} k_{y}\sigma_{x}+\alpha^{"} k_{y}\sigma_{z},
\end{equation}
where $H_{0}(k)$ is the part of Hamiltonian describing  the nearly-free-electron or -hole band dispersion written as 
\begin{equation}
\label{5}
H_{0}(k)= E_{0}(k)+ \delta_{1} k_{x}^{2} + \delta_{2} k_{y}^{2},  
\end{equation}
where $\delta_{1}$ and $\delta_{2}$ are related to the effective masses ($m_{x}^{*}$, $m_{x}^{*}$) by the relations $\left|\delta_{1}\right|=\hbar^{2}/2m_{x}^{*}$ and $\left|\delta_{2}\right|=\hbar^{2}/2m_{y}^{*}$, respectively. $\vec{\sigma}$ are the Pauli matrices describing spin degrees of freedom, while $\alpha$, $\alpha^{'}$, and $\alpha^{"}$ are the SOC parameters. 

Solving the eigenvalue problem involving the Hamiltonian of Eq. (\ref{4}), we obtain the following energy band dispersion,
\begin{equation}
\label{6}
E_{M}(k)= E_{0}(k)+ \delta_{1} k_{x}^{2} + \delta_{2} k_{y}^{2} +\pm \sqrt{{\alpha}^{2}k_{x}^{2}+({\alpha^{'}}^{2}+ {\alpha^{"}}^{2}) k_{y}^{2}}.  
\end{equation}
The corresponding eigenstates are given as
\begin{equation}
\label{7}
\psi_{k}^{\pm}=\frac{e^{i\vec{k}\cdot\vec{r}}}{\sqrt{2\pi(\rho_{\pm}^{2}+1)}} \begin{pmatrix}
\frac{i \alpha k_{x}-\alpha^{'} k_{y}}{\alpha^{"} k_{y}\mp \sqrt{{\alpha}^{2}k_{x}^{2}+({\alpha^{'}}^{2}+ {\alpha^{"}}^{2}) k_{y}^{2}} }\\
	1
\end{pmatrix},  
\end{equation}
where $\rho_{\pm}=\frac{{\alpha}^{2}k_{x}^{2}+{\alpha^{'}}^{2}k_{y}^{2}}{\left(\alpha^{"} k_{y} \mp \sqrt{{\alpha}^{2}k_{x}^{2}+({\alpha^{'}}^{2}+ {\alpha^{"}}^{2}) k_{y}^{2}}\right)^{2} }$. The expectation value of the spin operator is obtained from $\vec{s}^{\pm}=\frac{1}{2} \left\langle \psi_{k}^{\pm}\left|\vec{\sigma}\right|\psi_{k}^{\pm}\right\rangle$, resulting in
\begin{equation}
\label{8}
(S_{x}, S_{y}, S_{z})^{\pm}= \pm \frac{1}{2\sqrt{{\alpha}^{2}k_{x}^{2}+({\alpha^{'}}^{2}+ {\alpha^{"}}^{2}) k_{y}^{2}}}(\alpha^{'} k_{y}, \alpha k_{x}, \alpha^{"} k_{y}).
\end{equation}
As evident from Eq. (\ref{6}) that the anisotropic Rashba spin splitting is achieved depending on the magnitude of the SOC parameter in the spin-split bands along the $k_{x}$ ($M-\Gamma$) direction, $\alpha^{M-\Gamma}=\alpha$, and along the $k_{y}$ ($M-K$) direction, $\alpha^{M-K}=\sqrt{{\alpha^{'}}^{2} +{\alpha^{"}}^{2}}$. By fitting the DFT band dispersion of Fig. 3(a) along the $M-\Gamma$ and $M-K$ directions with Eq. (\ref{6}), we find that $\alpha^{M-\Gamma}=0.08$ eV\AA\ and $\alpha^{M-K}=1.34$ eV\AA. Since $\alpha^{M-\Gamma}<<\alpha^{M-K}$, the Rashba splitting exhibits a high degree of anisotropy, which is consistent with the spin-splitting energy shown in Fig. 3(b). Moreover, from Eq. (\ref{8}), we can estimate the ratio between $\alpha^{'}$ and $\alpha^{"}$ by comparing the in-plane and out-of-plane spin component along the $M-\Gamma$ direction, and found that $\alpha^{'}/\alpha^{"}=0.024$. Therefore, by using $\alpha^{M-K} =\sqrt{{\alpha^{'}}^{2}+ {\alpha^{"}}^{2}}=1.34$, we find that $\alpha^{'}=0.032$ eV\AA\ and $\alpha^{"}=1.33$  eV\AA. The larger value of $\alpha^{"}$ indicates that the spin-split states around the $M$ point are dominated by the out-of-plane spin component, which also agrees well with the calculated spin textures shown in Fig. 4(d).

Next, we discuss the origin of the anisotropic Rashba splitting around the $\Gamma$ point. The little point group of the wave vector associated with the $\Gamma$ point is $C_{3v}$, comprising of trivial identity operation ($E$), three-fold rotation $C_{3z}$, and three mirror planes ($M_{xz}$, $M_{xz}^{'}$, $M_{xz}^{"}$) [see Table II]. The effective $\vec{k}\cdot\vec{p}$ Hamiltonian for the ($k_{x}-k_{y}$) plane taking into account the symmetry invariants up to cubic in $k$ (see Table III) can be expressed as \cite{Vajna}
\begin{equation}
\label{9}
H_{\Gamma}(k)= H_{0}(k)+ \alpha \left(k_{y}\sigma_{x}- k_{x}\sigma_{y}\right) + \beta \left[ \left(k_{y}^{3}+k_{y} k_{x}^{2}\right)\sigma_{x} - \left({k_{y}}^{2} k_{x} + {k_{x}}^{3}\right) \sigma_{y}\right] + \gamma \left({k_{y}}^{3}-3(k_{y}k_{x}^{2})\right)\sigma_{z},
\end{equation}
where $\alpha$ is the linear term of the SOC parameter, while $\beta$ and $\gamma$ are the cubic terms of the SOC parameters. By defining $k=\sqrt{{k_{x}}^{2}+{k_{y}}^{2}}$ and $\phi=\arccos(k_{x}/k)$, where $\phi$ is the azimuthal angle of momentum $k$ with respect to the $x$-axis along the $\Gamma-M$ direction in the FBZ, we obtained that the Hamiltonian of Eq. (\ref{9}) can be written as
\begin{equation}
\label{10}
H_{\Gamma}(k)= H_{0}(k)+  \left( \alpha k + \beta k^{3} \right) \left(\sin(\phi) \sigma_{x} - \cos(\phi) \sigma_{y} \right) + \gamma k^{3} \sin(3\phi) \sigma_{z}.
\end{equation}
The energy eigenvalues of Eq. (\ref{10}) are given as:
\begin{equation}
\label{11}
E_{\Gamma}(k)= E_{0}(k)+ \delta_{1} k_{x}^{2} + \delta_{2} k_{y}^{2} +\pm \sqrt{  \left(\alpha k + \beta k^{3} \right)^{2}+ \gamma^{2}k^{6} \sin^{2}(3\phi)}.  
\end{equation}

We emphasized here that the second and third terms of $H_{\Gamma}(k)$ in Eq. (\ref{10}), respectively, will induce in-plane and out-of-plane spin polarization, so that the spin polarization at fixed energy is given by 
\begin{equation}
\label{12}
(S_{x}, S_{y}, S_{z})^{\pm}= \left[\pm \left( \alpha k + \beta k^{3} \right) \sin(\phi), \mp \left( \alpha k + \beta k^{3} \right) \cos(\phi), \mp \gamma k^{3} \sin(3\phi)\right],
\end{equation}
which is in agreement with the spin textures plots in Fig. 4(e). As demonstrated in Eq. (\ref{11}), the anisotropic Rashba spin splitting near the $\Gamma$ point is controlled by distinct parameters: $\alpha$, $\beta$, and $\gamma$, which vary based on the chosen $k$-path along the $\Gamma-M$ and $\Gamma-K$ orientations. Our fitting analysis of the spin-split bands illustrated in Figure 3(a) at the VBM, both along the $\Gamma-M$ and $\Gamma-K$ directions, utilizing Equation (\ref{11}), revealed the computed values for the SOC parameters: $\alpha^{\Gamma-M}=0.74$ eV\AA, $\alpha^{\Gamma-K}=0.88$ eV\AA, $\beta^{\Gamma-M}=52.8$ eV\AA$^{3}$, $\beta^{\Gamma-K}=37.8$ eV\AA$^{3}$, and $\gamma^{\Gamma-K}=0.9$ eV\AA$^{3}$. It's important to highlight that our fitting analysis of the spin-split bands along the $\Gamma-M$ direction yielded $\alpha^{\Gamma-M}=0$, as a result of the vanishing term $\gamma^{2}k^{6} \sin^{2}(3\phi)$ in Equation (\ref{11}) when the $k$-path is chosen along the $\Gamma-M$ direction. Consequently, the out-of-plane spin polarization $S_{z}$ along the $\Gamma-M$ direction becomes zero. This finding is consistent with the spin-resolved projected bands depicted in Figures 4(a)-(c) and the spin textures shown in Figure 4(e).

\begin{table*}
\caption{Several selected 2D materials systems supporting the anisotropic Rashba splitting. The SOC parameters are shown, including the linear terms [$\alpha^{M-\Gamma}$, $\alpha^{M-K}$ (in eV\AA)] of the SOC parameters around the $M$ point in the CBM and the linear terms [$\alpha^{\Gamma-M}$, $\alpha^{\Gamma-K}$ (in eV\AA)] and cubic terms [$\beta^{\Gamma-M}$, $\beta^{\Gamma-K}$, $\gamma^{\Gamma-M}$, $\gamma^{\Gamma-K}$ (all in eV\AA$^{3}$)] of the SOC parameters around the $\Gamma$ point in the VBM.} 
\begin{tabular}{cc  cc  cc   cc   cc  cc  cc  cc  cc} 
\hline\hline 
 2D Materials  &&  HSP  &&  $\alpha^{\Gamma-M}$ ($\alpha^{M-\Gamma}$) &&   $\alpha^{\Gamma-K}$ ($\alpha^{M-K}$)  && $\beta^{\Gamma-M}$  && $\beta^{\Gamma-K}$ &&   $\gamma^{\Gamma-M}$    &&   $\gamma^{\Gamma-K}$    &&  Ref. \\ 
\hline 
SbSi$_{2}$Bi     &&   $\Gamma$ ($M$)  &&  0.74 (0.08)     &&   0.88 (1.34)   &&  52.8   &&  37.8   &&0.0   &&  0.9    && This work \\SbGe$_{2}$Bi     &&   $\Gamma$ ($M$)  &&  1.45 (0.27)     &&   1.82 (1.45)   &&  63.9   &&  91.3   &&0.0   && -1.22   && This work \\  SbSn$_{2}$Bi     &&   $\Gamma$ ($M$)  &&  1.53 (0.29)     &&   1.95 (1.6)    && -212.8  && -280.1  &&0.0   && 108.3   && This work \\  BP               &&   $\Gamma$        &&  0.0109          &&   0.0036        &&       - &&      -  && -  && - && Ref. \cite{Popovi}\\	
$M$SSe ($M$=Mo,W)&&   $\Gamma$        &&  -            &&   0.004 - 0.17    &&  - &&  -    &&  -    &&  - && Ref. \cite{Chakraborty}\\ 
Mo$XY$ ($X,Y$= S,Se, Te)           &&   $\Gamma$        &&  0.077 - 0.479  &&0.77 - 0.487  &&  - &&  - &&  -    &&  -      && Ref. \cite{Hu}\\ 
W$XY$ ($X,Y$= S,Se, Te)            &&   $\Gamma$        &&  0.157 - 0.524  && 0.158 - 0.514&&  -    &&  -    &&  -    &&  -     && Ref. \cite{Hu}   \\
Pt$XY$ ($X,Y$= S,Se, Te)            &&   $\Gamma$        && 0.435-1.654  && 0.746 - 1.333&&  -    &&  -    &&  -    &&  -      && Ref. \cite{Sino}   \\
$M$Si$_{2}$P$_{x}$As$_{y}$ ($M$= W, Mo)            &&   $\Gamma$        && 0.0 - 0.6  && 0.073 - 2.77 &&  -    &&  -    &&  -    &&  -      && Ref. \cite{Rezavand}   \\
BiTeI            &&   $\Gamma$        && 1.82  && -   &&  -    &&  -    &&  -    &&  -      && Ref. \cite{Zhang9}   \\
\hline\hline 
\end{tabular}
\label{table:Table 4} 
\end{table*}

We summarize the anisotropic Rashba parameters of the 2D Janus $XA_{2}Y$ MLs around the $M$ and $\Gamma$ point in the CBM and VBM, respectively, in Table IV and compare the result with a few selected 2D materials supporting anisotropic Rashba splitting from previously reported data. In particularly, the magnitude of the linear terms ($\alpha^{M-\Gamma}$, $\alpha^{M-K}$, $\alpha^{\Gamma-M}$, $\alpha^{\Gamma-K}$) of the SOC parameters are much larger than that observed on black-phosphorene \cite{Popovi}, and 2D Janus $M$Si$_{2}$P$_{x}$As$_{y}$ MLs \cite{Rezavand}, and 2D Janus TMDCs MLs including $M$SSe Mls \cite{Chakraborty}, $MXY$ MLs \cite{Hu}, and Pt$XY$ MLs \cite{Sino}, and are comparable with that observed on BiTeI ML \cite{Zhang9}. Notably, the magnitudes of these SOC parameters increase as the Janus $XA_{2}Y$ MLs become heavier. For instance, the $\alpha^{M-K}$ value around the $M$ point in the Janus SbSn$_{2}$Bi ML (1.6 eV\AA) far exceeds that of the Janus SbSi$_{2}$Bi ML (1.34 eV\AA), as shown in Table IV. To enhance the spin splitting magnitude in these materials, it's possible to introduce heavier elements with strong SOC through doping \cite{Volobuev} and apply strain \cite{Anshory, Arras}, making the experimental detection of this phenomenon is more feasible, particularly utilizing techniques like spin-resolved photoemission spectroscopy.

\begin{figure*}
	\centering
		\includegraphics[width=1.0\textwidth]{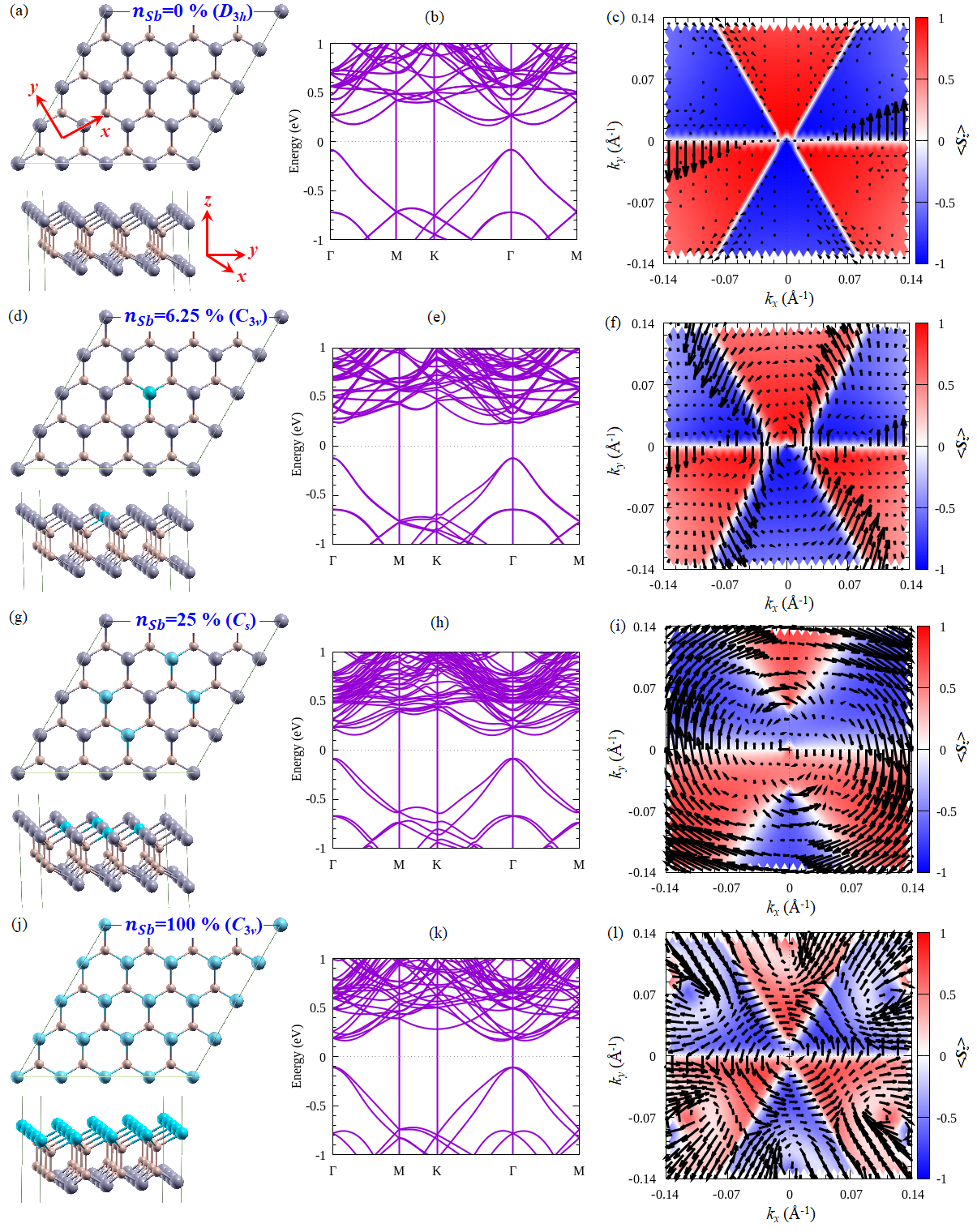}
	\caption{The crystal structure in the top and side views, spin-split bands, and spin textures for Si$_{2}$Bi$_{2}$ ML supercell with surface imperfection are shown for: (a)-(c) the clean system and doping systems with (d)-(f) Sb concentration $n_{\texttt{Sb}} =6.25$\%, (g)-(i) with Sb concentration $n_{\texttt{Sb}} =25$\%, and (j)-(l) Sb concentration $n_{\texttt{Sb}} =100$\%.}
	\label{fig:Figure5}
\end{figure*}

We have discovered that the Janus $XA_{2}Y$ MLs exhibit large and anisotropic Rashba splitting, suggesting their potential utility in spintronics applications. Nonetheless, the creation of these Janus $XA_{2}Y$ MLs is expected to be sensitive to surface imperfections, akin to what has been observed in previous Janus TMDCs MLs. In typical TMDCs, the Janus arrangement is achieved through techniques involving sulfurization \cite{Z_Jing, Sant2020} and selenization \cite{Lu2017}, where one plane of the ML surface consists of different chalcogen atoms. Moreover, recent experimental work has demonstrated notable advancements in controlling the chalcogen exchange processes \cite{Z_Jing, Sant2020, Lu2017, Lin2020, Qin, Jang2022}, enabling the formation of the surface imperfection through selective modification of only a fraction of the chalcogen atoms in one of the surface layers of TMDCs MLs. By employing a similar principle, it is also feasible to create Janus $XA_{2}Y$ MLs through surface imperfections by partially substituting the group V ($X$, $Y$) atoms in one of the surface layers of pure $A_{2}X_{2}$ ($A_{2}Y_{2}$) MLs. Recently, 2D mono elemental (Bi and Sb) MLs with a buckled honeycomb lattice structure containing two layers have been successfully synthesized experimentally \cite{Jian, WuXu}. Hence, there exists an opportunity to experimentally synthesize MLs containing $A$Bi ($A$Sb) by substituting the Bi (Sb) layer on one side with an $A$ (Si, Ge, Sn) layer. Assuming successful experimental synthesis of SiBi MLs, it becomes feasible to connect two SiBi MLs in a Bi-Si-Si-Bi configuration, potentially leading to the creation of a pure Si$_{2}$Bi$_{2}$ ML comprising four atomic layers. The surface imperfection of the Si$_{2}$Bi$_{2}$ ML can be further achieved by interstitially substituting the Sb atom in the Bi site within one of the surface layers. The fully Janus SbSi$_{2}$Bi ML structure can be realized when all the Bi site atoms in one of the surface layers are completely replaced by Sb atoms. Indeed, the favorable formation of the complete Janus SbSi$_{2}$Bi ML is confirmed by the reduced formation energy as the Sb concentration increases, as shown in Table S1 in the Supplementary Materials \cite{Supporting}. We would like to note that the formation of the Janus SbSi$_{2}$Bi ML is more favorable when achieved through the surface imperfection of the Si$_{2}$Sb$_{2}$ ML due to the lower formation energy; see Table S1 in the Supplementary Materials \cite{Supporting}.

To ensue discussion aiming to explore the pivotal role of the surface imperfections in influencing the spin-splitting characteristics of Janus $XA_{2}Y$ MLs, we show in Figure 5 the relation between crystal symmetry, spin-split bands, and spin textures around the $\Gamma$ point at the VBM of the Si$_{2}$Bi$_{2}$ ML under different concentration of the Sb substitutions. The Sb atom concentration ($n_{\texttt{Sb}}$) is expressed as a percentage relative to the number of Bi atoms on the upper surface within the framework of the 4x4x1 supercell of the Si$_{2}$Bi$_{2}$ ML. When $n_{\texttt{Sb}}=0$\%, representing the pure Si$_{2}$Bi$_{2}$ ML, the crystal's $D_{3h}$ point group symmetry [Fig. 5(a)] enforces the appearance of the spin splitting in the electronic bands, except for $k$ bands along the $\Gamma-M$ line [Fig. 5(b)], leading to a fully out-of-plane spin polarization with a three-fold pattern [Fig. 5(c)]. With the introduction of a single Sb atom interstitially replacing the Bi atom within one of the surface layer of the Si$_{2}$Bi$_{2}$ ML ($n_{\texttt{Sb}}=6.25$\%) [see Fig. 5(d)], the in-plane mirror symmetry $M_{xy}$ is broken, resulting in the crystal's $C_{3v}$ point group symmetry. This change lifts the spin degeneracy of the bands along the $\Gamma-M$ line [Fig. 5(e)], giving rise to a three-fold pattern in both in-plane and out-of-plane spin polarization [Fig. 5(f)]. Interestingly, as the Sb concentration is increased to $n_{\texttt{Sb}}=25$\% [Fig. 5(g)], the in-plane mirror symmetry $M_{xy}$ and the three-fold rotation symmetry $C_{3z}$ are both disrupted, yielding a stable configuration with $C_{s}$ point group symmetry. Consequently, the spin-splitting bands exhibit pronounced anisotropy [Fig. 5(h)], showcasing a reverse pattern of out-of-plane spin polarization mirrored at $k_{x}=0$. Finally, with a Sb substitution of $n_{\texttt{Sb}}=100$\%, the Janus SbSi$_{2}$Bi ML forms [Fig. 5(j)], featuring the crystal's $C_{3v}$ point group symmetry. This configuration shares the spin-split band and spin polarization characteristics seen in the $n=6.25$\% scenario but with a notably higher spin splitting energy [Figs. 5(k)-5(l)].

\begin{table*}
\caption{The dependent of the SOC parameters on the concentration and configuration of the Sb atom on the top surface of Si$_{2}$Bi$_{2}$ ML modelled by $4\times4\times1$ supercell. The Sb atom concentration ($n_{\texttt{Sb}}$) is expressed as a percentage relative to the number of Bi atoms on the upper surface within the framework of the $4\times4\times1$ supercell. Concentration of $n_{\texttt{Sb}}=0$\% means that the system is pure Si$_{2}$Bi$_{2}$ ML, while the concentration of $n_{\texttt{Sb}}=100$\% indicates that the system forms the Janus structure of SbSi$_{2}$Bi ML. The SOC parameters including the linear terms [$\alpha^{\Gamma-M}$, $\alpha^{\Gamma-K}$ (all in eV\AA)] and cubic terms [$\beta^{\Gamma-M}$, $\beta^{\Gamma-K}$, $\gamma^{\Gamma-M}$, $\gamma^{\Gamma-K}$  (all in eV\AA$^{3}$)] of the SOC parameters are shown for the VBM around the $\Gamma$ point.} 
\begin{tabular}{ccc ccc ccc  ccc   ccc   ccc  ccc ccc  ccc  } 
\hline\hline 
 Concentration (\%)  &&&  PGS  &&&   HSP  &&&  $\alpha^{\Gamma-M}$     &&&   $\alpha^{\Gamma-K}$  &&& $\beta^{\Gamma-M}$  &&& $\beta^{\Gamma-K}$ &&&  $\gamma^{\Gamma-M}$    &&& $\gamma^{\Gamma-K}$ \\ 
\hline 
 0         &&&   $D_{3h}$  &&&   $\Gamma$         &&&  -        &&&   -      &&& -          &&& -    &&& -   &&&  0.32     \\
 6.25      &&&   $C_{3v}$  &&&   $\Gamma$         &&&  0.09     &&&   0.01   &&&  -70.0     &&& 79.9 &&& 0.0   &&&  0.4    \\
 12.5      &&&   $C_{s}$   &&&   $\Gamma$         &&&  0.11     &&&   0.17   &&&  -         &&& -     &&& -   &&&  -     \\
  25      &&&    $C_{s}$   &&&   $\Gamma$         &&&  0.34     &&&   0.37   &&&  -         &&& -     &&& -   &&&  -     \\
	50      &&&    $C_{s}$   &&&   $\Gamma$         &&&  0.21     &&&   0.36   &&&  -         &&& -     &&& -   &&&  -     \\
	75      &&&    $C_{s}$   &&&   $\Gamma$         &&&  0.25     &&&   0.42   &&&  -         &&& -     &&& -   &&&  -     \\
 100      &&&    $C_{3v}$  &&&   $\Gamma$         &&&  0.74     &&&   0.88   &&&  52.8      &&& 52.8  &&& 0.0  &&&  0.9   \\
\hline\hline 
\end{tabular}
\label{table:Table 5} 
\end{table*}

The impact of Sb atom substitution on the spin-splitting characteristics of the Si$_{2}$Bi$_{2}$ ML is further quantified by evaluating the SOC parameters through the application of a symmetry-adapted SOC Hamiltonian for the $D_{3h}$, $C_{3v}$, and $C_{s}$ point groups; see the Supplementary Materials for the explicit form of the $\vec{k}\cdot\vec{p}$ model \cite{Supporting}. The computed results for these SOC parameters are detailed in Table V. Notably, the linear terms of the SOC parameters experience significant enhancement with the progressive substitution of Sb atoms. This suggests that the partial substitutions of the group V ($X, Y$) atoms on the pure $A_{2}X_{2}$ ($A_{2}Y_{2}$) MLs offer an effective means to adjust the spin-splitting characteristics of Janus $XA_{2}Y$ MLs. Thus, our discoveries highlight the potential spintronic capabilities of Janus $XA_{2}Y$ MLs, with experimental realization becoming feasible.

Before closing, we would like to discuss the potential application of the current system in spintronics devices, specifically SFETs \cite{Datta}. In SFETs, the response of the Rashba parameter to the external electric field, measured as the electric field response rate of the Rashba parameter denoted as $\left|\Delta \alpha/\Delta|\vec{E}|\right|$, plays a crucial role in their operation. In fact, the strong response of the Rashba parameter to the external electric field has been previously reported on Bi-doped ZnO nanowires \cite{Aras2018}. To maintain spin coherence in SFETs, it is essential to have a large Rashba parameter and a strong response to the applied electric field, as this helps reduce the length of the spin channel. By examining the Rashba parameter of the Janus SbSi$_{2}$Bi ML near the $\Gamma$ point at the VBM, we have revealed that it exhibits a notably linear response to the electric field, as depicted in Figure S2 of the Supplementary Materials \cite{Supporting}. This response results in a significant response rate $\left|\Delta \alpha/\Delta|\vec{E}|\right|$ of up to 0.59 e\AA$^{2}$ and 0.44 e\AA$^{2}$ along the $\Gamma-K$ and $\Gamma-M$ directions, respectively. These response rates are comparable to that reported previously on the typical 2D Rashba systems including BiSb (0.92 e\AA$^{2}$) \cite{Wu2021}, $T$-RbPb$X_{3}$ ($X$= Br, I) (0.177 - 0.544 e\AA$^{2}$) \cite{Chen2021}, and TlSn$X_{3}$ ($X$= Br, I) (0.23 - 0.79 e\AA$^{2}$) \cite{Jin}. Considering the Janus SbSi$_{2}$Bi ML as the quasi-one-dimensional channel of the SFET, electron spins can undergo precession due to the Rashba effect, characterized by the precession angle $\theta=2\alpha m^{*}L/\hbar^{2}$, where $L$ represents the channel length of the SFET \cite{Supporting}. By adjusting the gate voltage of the SFET, we can manipulate the Rashba parameter $\alpha$, thereby controlling the precession angle $\theta$. Thanks to the substantially large $\left|\Delta \alpha/\Delta|\vec{E}|\right|$ exhibited by the SbSi$_{2}$Bi ML, we have found that the estimated channel length $L$ is approximately 159 nm \cite{Supporting}). This length is comparable to the spin channel length of other primitive systems for SFETs such as BiSb ($L = 158$ nm) \cite{Wu2021}, $T$-RbPb$X_{3}$ ($X$= Br, I) ($L = 72 - 172$ nm) \cite{Chen2021}, and TlSn$X_{3}$ ($X$= Br, I) ($L = 102 - 220$ nm) \cite{Jin}, but is significantly smaller than the typical values of conventional SFETs ($L\approx 2-5$ $\mu$m) operating at room temperature \cite{Takase2017, Chuang2015, Trier2020}. Remarkably, the Janus $XA_{2}Y$ ML with large Rashba parameters and strong electric field response rates could serve as promising candidates for SFETs.

\section{Conclussion}

In summary, we employed first-principles density-functional theory (DFT) calculations coupled with $\vec{k}\cdot\vec{p}$-based symmetry analysis to systematically explore the properties related to SOC in the 2D Janus $XA_{2}Y$ monolayer (ML) composed of group IV-V compounds. The absence of in-plane mirror symmetry in these Janus $XA_{2}Y$ MLs reduces the crystal's point group symmetry, causing the Rashba SOC to exhibit anisotropy around specific high symmetry $\vec{k}$ points within the FBZ. This anisotropy also maintains pronounced out-of-plane spin patterns. We have focused our examination on the Janus SbSi$_{2}$Bi ML as a representative case, where these properties are particularly noticeable near the CBM around $M$ point and the VBM around $\Gamma$ point. Additionally, our investigation uncovered that the anisotropic spin splitting and out-of-plane spin patterns are notably influenced by surface imperfections. This dependence on imperfection characteristics encompasses the composition and configuration of $X$ and $Y$ elements within the 2D Janus $XA_{2}Y$ MLs.

Considering the anisotropic Rashba phenomenon observed in our study, characterized by predominant out-of-plane spin polarization, it is reasonable to anticipate that this effect can be replicated in other 2D materials possessing a trigonal prismatic structure. The asymmetry in the in-plane mirror operation $M_{xy}$ is solely responsible for driving this phenomenon. Our symmetry analysis has elucidated that various 2D systems exhibit the potential for hosting the anisotropic Rashba effect with predominant out-of-plane spin polarization. These systems include 2D Janus structures of group V-IV-III-VI MLs \cite{Lin}, MLs of group III-V compounds \cite{Mustafa}, and 2D Janus group III monochalcogenides MLs with oxygenation \cite{Vu}. As a result, our projections are anticipated to stimulate further exploration through theoretical and experimental investigations, aiming to discover new 2D materials that support the anisotropic Rashba effect. Such discoveries hold promise for advancing spintronic applications in the future.

\begin{acknowledgments}

This work was supported by the Rekognisi Tugas Akhir (RTA) program (No. 5075/UN1.P.II/Dit-Lit/PT.01.01/2023) supported by Gadjah Mada University, Indonesia. The first author (AL) would like to thank the Indonesian Endowment Fund for Education (LPDP) Indonesia for financial support through the LPDP scholarship program. The computation in this research was performed using the computer facilities at Gadjah Mada University, Indonesia. 

\end{acknowledgments}

\bibliography{Reference1}


\end{document}